\documentclass[final,5p,times,twocolumn]{elsarticle}

\usepackage{graphicx}
\usepackage[english]{babel}
\usepackage[utf8]{inputenc}
\usepackage[T1]{fontenc}
\usepackage{hyperref}

\usepackage{amssymb}
\usepackage{amsfonts}
\usepackage{amsmath}
\usepackage{color}
\usepackage{soul}

\usepackage{color}

\newcommand{\textitb}[1]{{\color{black} #1}}

\usepackage{upgreek}

\newcommand{\hz}{\mathcal{H}_0}
\newcommand{\hp}{H_{0\textrm{P}}}
\newcommand{\hsn}{H_{0\textrm{SN}}}

\newcommand{\ompl}{\Omega_{0m}^{\textrm{P}}}
\newcommand{\omsn}{\Omega_{0m}^{\textrm{SN}}}

\journal{}

\begin{document}

\begin{frontmatter}



\title{Slow-rolling scalar dynamics as solution for the Hubble tension}


\author[a1,a2]{Giovanni Montani}
\author[a1]{Nakia Carlevaro}
\author[a3,a4,a5]{Maria Giovanna Dainotti}

\address[a1]{ENEA, Nuclear Department, C.R. Frascati, Via E. Fermi 45, 00044 Frascati (Roma), Italy}
\address[a2]{Physics Department, ``Sapienza'' University of Rome, P.le Aldo Moro 5, 00185 Roma, Italy}
\address[a3]{Division of Science, National Astronomical Observatory of Japan, 2-21-1 Osawa, Mitaka, Tokyo 181-8588, Japan}
\address[a4]{The Graduate University for Advanced Studies (SOKENDAI), Shonankokusaimura, Hayama, Miura District, Kanagawa 240-0115}
\address[a5]{Space Science Institute, Boulder, CO, USA}

\begin{abstract}
We construct a theoretical framework to interpret the Hubble tension
by means of a slow-rolling dynamics of a self-interacting scalar field.
In particular, we split the Friedmann equation in order to construct a
system for the three unknowns,
corresponding to the Hubble parameter $H$, the scalar field $\phi$ and its
self-interaction potential $V$, as functions of the redshift. 
In the resulting picture, the vacuum energy density is provided by a
constant term in the potential $V(\phi)$, while the corresponding small kinetic term is responsible for reproducing the apparent variation of
the Hubble constant $H_0$ with the redshift. 
The emerging solution depends on two free parameters, one of which is
fixed to account for the discrepancy between the values of $H_0$ as
measured by the Super Nova Ia sample ($H_0=73.6\pm1.1$ km s$^{-1}$ Mpc$^{-1}$ (Brout et al., 2022)) and the Planck satellite data ($H_0=67.4\pm0.5$ km s$^{-1}$ Mpc$^{-1}$ (Aghanim et al., 2020)),
respectively. The other parameter is instead determined by a fitting
procedure of the apparent Hubble constant variation across the data
corresponding to a 40 bin analysis of the Super Nova Pantheon sample, in each of
which $H_0$ has been independently determined.
The fundamental result of the present analysis is the emerging Hubble
parameter as function of the redshift, which correctly takes the Super Nova Ia
prediction at $z=0$ and naturally approaches the profile predicted by a
flat $\Lambda$CDM model corresponding to the cosmological parameters
detected by Planck. It is remarkable that this achievement is reached without reducing the Super Nova Ia data to a single point for determining $H(z=0)$, but accounting for the distribution over their redshift interval of observation,
via the binned analysis.
\end{abstract}

%\begin{keyword}
%key1 \sep
%key2 \sep
%key3
%\end{keyword}

%\pacs{97.10.Gz; 95.30.Qd}
%% PACS Nos.: 52.30.-q; 95.30.Qd; 97.10.Gz

\end{frontmatter}

% \linenumbers

\section{Introduction}
\thispagestyle{empty}
The most relevant puzzle emerged 
in recent years in Precision Cosmology is the (4.9 $\sigma$) inconsistency between the detection of the value 
of the Hubble constant $H_0$ from 
near sources, like the Pantheon and 
the Pantheon+ samples of the Type Ia Supernovae (SNe Ia) \cite{Brout:2022vxf,Scolnic_2022,Pan-STARRS1:2017jku}, and from the Cosmic Microwave Background (CMB) data, taken by the Planck Satellite \cite{Planck:2018vyg}:
this discrepancy is commonly dubbed \emph{Hubble tension} \cite{DiValentino:2021izs}. In this respect, for a data analysis which reduces the 
error bars for the Pantheon+ sample, 
so enforcing the tension, see \cite{Dainotti:2023ebr,2022ApJ...938...36R}; example of new physics approach, in addition to the dark energy Universe component, can be found instead in \cite{2020PhRvD.102b3518V,2023Univ....9..393V}. See also \cite{arXiv-2402-04767}, for an analysis of the tension associated to the Lambda Cold Dark Matter ($\Lambda$CDM) models.

An important issue in the understanding of the Hubble tension came out 
in \cite{apj-powerlaw}, where it was shown that, 
for the SN Ia Pantheon sample, the value 
of $H_0$ does not remain constant within $2\sigma$ confidence level. 
The obtained effective profile $H_0\to\hz(z)$ emerged from a binned analysis, which separated the total sample into equipopulated redshift (dubbed $z$) intervals. 
For an extension of this analysis, 
which makes allowance for varying 
the cosmological parameters, see \cite{galaxies10010024}. 
We here stress that this analysis has been performed with the assumption of Gaussian likelihoods, as commonly assumed, however a more recent treatment of the data shows the non-Gaussianity of the likelihood of the SNe Ia, for details about this topic see \citep{DAINOTTI202430,Dainotti2023ApJ...951...63D,Bargiacchi2023MNRAS.521.3909B}.
Furthermore, correlated analyses providing information about the non-constant character of 
$H_0$ over the SN Ia samples can be found in \cite{Krishnan:2020vaf,Krishnan:2020obg,Malekjani:2023dky,kazantzidis}. The specific behavior determined in \cite{apj-powerlaw} was of the form $\hz(z)=H_0(1+z)^{-\alpha}$, where $\alpha$ is 
a parameter of the order of magnitude $10^{-2}$. A model to provide a theoretical explanation to such a decaying 
behavior has been developed in \cite{schiavone_mnras}. This implements a metric $f(R)$-gravity in 
the Jordan frame \cite{Sotiriou-Faraoni:2010,NOJIRI201159}, using 
the emerging non-minimally coupled 
scalar field to account for the $H_0$ 
variation with the redshift. 
A similar study, which however 
includes dynamical dark energy in the 
system evolution \cite{DiValentino:2020naf}, can be found in \cite{deangelis-fr-mnras} (there, $\hz$ rapidly decays and 
the Hubble tension disappears already 
at redshift $\simeq 5$ - up to $z\simeq 2$ when accounting for the errors). 
For other approaches to provide an explanation for 
the Hubble tension in $f(R)$-gravity 
see \cite{Nojiri:2022ski,Odintsov:2020qzd} and studies of 
the role that the local Universe inhomogeneities could play in the tension solution can be found in \cite{apj-powerlaw}.

Here, we propose a different theoretical scenario, based on an external 
minimally coupled scalar field which, 
performing a slow-rolling dynamics, is responsible, on one hand for reproducing a $\Lambda$CDM model and, on the other hand for rescaling the value of the Planck Hubble constant by a slow varying function with the redshift. 
More specifically, the scalar field 
potential has a dominant constant component, which is the source of the present day vacuum energy density, while 
the corresponding kinetic term induces 
the requested variation $\hz$. \textitb{This self-interacting scalar field is then responsible for a quintessence slow-rolling, having nothing to do, in principle, with the early Universe inflation.} The final output of the model is a 
rescaling of the Planck $H_0$ value, which is reconciled with the SN Ia measurement at present days.
Such $\hz$ is different with respect to the profile predicted in \cite{apj-powerlaw}, but it possesses the same essential features and phenomenological implications.

Using data associated to the binned analysis of the SN Ia Pantheon sample, we perform a fitting procedure to determine one of the two free parameters of the model, while the other one is determined by reproducing the required ratio between the $H_0$ values measured by SN Ia and by Planck. We also provide a comparison of the present model with the behavior $(1+z)^{-\alpha}$. The comparison of the two models is rather immediate and it turns out a statistic equivalence. However, the slow-rolling model here proposed appears a simpler solution to the Hubble tension than the $f(R)$ paradigm developed in \cite{schiavone_mnras,apj-powerlaw}, and comparable in its capability to fit data with respect to the profile $(1+z)^{-\alpha}$.

Finally, having fixed both the free parameters of the model, we plot
the Hubble parameter $H(z)$, which correctly takes in $z=0$ the value of
the Hubble constant detected via SNe Ia and it also overlaps
for larger redshift (already for $z\gtrsim 5$) the profile that is
associated to a flat $\Lambda$CDM model specified for the Planck Satellite measured parameters. The remarkable feature of the comparison of our model with the data is
that the information coming from the SN Ia sample is not compressed in
providing only the value $H_0$, but their distribution over the
redshift is taken into account via the 40 bin data set we use in
constraining one free parameter. In this respect, we stress that, by construction, $H(z)$ properly fits all the 40 points of the bin redshift distribution. We conclude by underlining how the
proposed scenario combines together two
valuable and novel aspects, i.e.
the natural (phantom free) scalar matter proposal and a more detail comparison of the information contained in SN Ia Pantheon sample and the Planck data.

The paper is organized as follows. In Sec. \ref{sec2}, we address the Friedmann equation for a flat geometry by including in the dynamics a self interacting scalar field requiring its slow rolling behavior. In Sec. \ref{sec3}, the obtained dynamical system in solved by means of specific assumptions. We determine an expression of the Hubble function depending on two free parameters. One of them is directly set by assuming that the present day Hubble parameter $H(z=0)$ corresponds to $H_0$ measured by the SN Ia sample. Finally, we express $H$ in terms of a flat $\Lambda$CDM model but with an effective Hubble constant which is now a function of the redshift. In Sec. \ref{sec4}, using the 40 bin data set we perform a fit procedure for the remaining free parameter and we compare the obtained behavior with a power law scenario. Then we plot the Hubble parameter $H(z)$ to elucidate how our model is a reliable candidate to solve the Hubble tension. Concluding remarks follow.

\section{Slow-rolling dynamics}\label{sec2}

We consider a flat Robertson-Walker geometry (accordingly to experimental constraints \cite{2020MNRAS.496L..91E}), having the line element
\begin{equation}
	ds^2 = - dt^2 + a^2(t) \delta_{ij}dx^idx^j
	\, , 
	\label{ioma1}
	\end{equation}
where $(i,j)=1,2,3$, $t$ denotes the synchronous time (we are in $c=1$ units) and $a(t)$ is the cosmic scale factor, regulating the Universe expansion. We include into the dynamics a self-interacting scalar field $\phi(t)$, characterized by the potential $V(\phi)$ and a standard matter contribution $\rho_m = \rho_{0m}(a_0/a)^3$, where $\rho_{0m}$ and $a_0$ denote the present day value of the matter energy density and of the scale factor, respectively. \textitb{We stress that, throughout this paper, the present day value of a quantity is denoted by the suffix 0.} The Friedmann equation ($00$-component of the Einstein  equations) thus takes the following form:
\begin{equation}
	H^2(t) \equiv \left( \frac{\dot{a}}{a(t)}\right)^2 = \frac{\chi}{3}\left( 
	\frac{\rho_{0m}a_0^3}{a^3} + \frac{1}{2}\dot{\phi}^2 + V(\phi )\right)
	\, , 
	\label{ioma2}
\end{equation}
where the dot indicates time differentiation, $H(t)$ is the Hubble parameter and $\chi$ denotes the Einstein constant. On the space-time described by the line element ion Eq.(\ref{ioma1}), the scalar field dynamics is provided by the 
Klein-Gordon-like equation
\begin{equation}
	\ddot{\phi} + 3 H \dot{\phi} + 
	\frac{dV}{d\phi} = 0
	\, .
	\label{ioma3}
\end{equation}

The Universe dynamics is summarized by the two degrees of freedom $a(t)$ and $\phi(t)$ and we now implement in Eqs.(\ref{ioma2}) and (\ref{ioma3}) the requests to deal with a ``slow-rolling behavior'' of the scalar field, i.e. for $\phi$ in a given interval, we assume the following conditions:
\begin{equation}
	V(\phi ) = \rho_{\Lambda} + G(\phi )\;,\qquad\dot{\phi}^2/2 \ll V(\phi )\;,\qquad
 \ddot{\phi}\ll 3H\dot{\phi}\;, 
	\label{ioma4}
\end{equation}
where $\rho_{\Lambda}$ denotes a constant energy density.
Here $G(\phi)$ is a generic correction to a flat self-interaction potential to be determined via the model 
details. It is useful to stress that, 
in what follows, independently of their smallness, in Eq.(\ref{ioma2}) we retain  both this potential corrections, as well as, the kinetic term, while in Eq.(\ref{ioma3}) we neglect the second time derivative of the scalar field. We thus rewrite Eqs.(\ref{ioma2}) and (\ref{ioma3}) as
\begin{align}
	H^2 = \frac{\chi}{3}\left( 
	\frac{\rho_{0m} a_0^3}{a^3} + \rho_{\Lambda}\right) + \frac{\chi}{3}\left(\frac{1}{2}\dot{\phi}^2 + G(\phi )\right)
\,, \label{ioma5}\\
	3H \dot{\phi} + \frac{dG}{d\phi} = 0
	\, , 
	\label{ioma6}
\end{align}
respectively.

In the following, we analyze the two equations above having in mind the construction of a $\Lambda$CDM model, allowing for an explanation of the Hubble tension.

\section{Paradigm for the Hubble tension}\label{sec3}

\textitb{Let us describe the dynamics of the system in terms of the redshift variable $z(t)$ defined as $1 + z(t) = a_0/a(t)$. We thus obtain the relation  $(...)\dot{\,} =-H(1+z)(...)'$, where the prime indicates derivative with respect to $z$, and the dynamical equation for the scalar field, i.e. Eq.(\ref{ioma6}), can be restated as
\begin{align}
3H^2 (1+z) \phi'^2=G'\,.\label{new6}
\end{align}

In order to construct three equations for the three unknowns $H(z)$, $\phi(z)$, $G(z)$ (from the latter two, we can calculate $G(\phi)$), we require that the following condition holds:
\begin{equation}
H^2(1+z)^2\phi'^2=-\beta\, G(\phi(z))\, , \label{ioma8}
\end{equation}
where $\beta$ is a free parameter of the model. In this framework, the system of Eqs.(\ref{ioma8}) and (\ref{new6}) can be solved obtaining the functional form of $G(z)$:
\begin{align}\label{gform}
G(x)= G_0 (1+z)^{-3\beta}\;.%\qquad G_0=-H_0^2\phi_0'^2/\beta\;,
\end{align}
%where the $G_0$ constant .
%we emphasize that $H_0$ here is a constant value of the present day ($z=t=0$) Hubble parameter $H(0)$.
This solution formally completes the model dynamics once substituted into the Friedman equation, i.e. Eq.(\ref{ioma5}), providing the evolution equation for $H(z)$.
}

Let us introduce a normalizing fiduciary constant $H_*$. Using the expressions above, Eq.(\ref{ioma5}) can be finally recast as
\begin{align}\label{Hform}
H(z) = H_*\; \sqrt{\Omega_{m}^{*} (1+z)^{3}+\Omega_{\Lambda}^{*}+\Omega_{\phi}^{*} (1+z)^{-3\beta}}\;,
\end{align}
where we have defined the dimensionless constants
\begin{align}\label{omegas}
\Omega_{m}^{*} \equiv \frac{\chi\rho_{0m}}{3H_*^2}\;,\quad
\Omega_{\Lambda}^{*} \equiv \frac{\chi \rho_{\Lambda}}{3H_*^2}\;,\quad
\Omega_{\phi}^{*} \equiv \frac{\chi G_0}{3H_*^2}\Big(1-\frac{\beta}{2}\Big)\;.
\end{align}
\textitb{We stress that, using such a peculiar normalization for Eq.(\ref{ioma5}), $(\Omega_{m}^{*}+\Omega_{\Lambda}^{*}+\Omega_{\phi}^{*})$ results equal to one only if $H_*$ coincides with $H_0$, i.e. the present day ($z=t=0$) value of the Hubble parameter $H(0)$}. Here $\Omega_{m}^{*}$ and $\Omega_{\Lambda}^{*}$ denote the matter and vacuum energy critical parameters, respectively. $\Omega_{\phi}^{*}$ represents instead the corresponding contribution associated to the scalar field. We also stress that the consistency of the model requires the following relation:  sign$(\Omega_{\phi}^{*})$=sign$(1/2-1/\beta)$. In fact, the form of the scalar field $\phi(z)$ can be derived from its evolution equation: by using Eqs.(\ref{new6}), (\ref{gform}) and the expression of $\Omega_{\phi}^*$ in Eq.(\ref{omegas}), we obtain
\begin{align}\label{phievo}
\bar{\phi}'=\sqrt{\frac{H_*^2 \;\Omega_{\phi}^*\;(1+z)^{-3\beta}}{(1/2-1/\beta)\;H(z)^2}}\;,
\end{align}
where we have defined the dimensionless potential $\bar{\phi}=\phi\sqrt{\chi/3}$.

The Hubble parameter evolution in Eq.(\ref{Hform}) can be now specified for measured quantities having in mind a flat $\Lambda$CDM model. We introduce the following notation concerning the values of the Hubble constant $H_0$ via different sources: we denote by $\hp$ the measurement from the CMB by the Planck Satellite, while $\hsn$ indicates the value detected from the SNe Ia. Here, such quantities are taken as derived from a flat $\Lambda$CDM model with $\ompl$ and $\omsn$, respectively. In order to overcome the discrepancy between the detected values of the present day Hubble constant, let us now impose that $H_0$ is obtained by the SNe Ia sample, and, at the same time, that the appropriate early time Universe behavior is recovered, i.e. we set 
\begin{align}\label{H0expression}
H_0=\hsn\;,\qquad
H_*=\hp\;.
\end{align}
\textitb{We point out that the early Universe evolution can be obtained only if $\beta>-1$, in order to neglect the contribution of $\Omega_\phi^*$ at large $z$. In this scheme, we obtain}
\begin{align}
\Omega_m^*&=\ompl\;,\qquad
\Omega_\Lambda^*=(1-\ompl)\;,\label{planck-omega}\\
\Omega_\phi^*&=\Omega_{0\phi}\equiv \chi G_0(1-\beta/2)/(3\hp^2)\;.     
\end{align}
Moreover, following the consideration above, at $z=0$ we require the relation
\begin{align}\label{omegaphi}
\ompl + (1-\ompl) +\Omega_{0\phi} = \hsn^2/\hp^2\;,
\end{align}
which formally defines $\Omega_{0\phi}$ and we recall that $\hsn>\hp$ (for a similar approach, see the discussion in \cite{2021MNRAS.505.3866E}). \textitb{Once $\Omega_{0\phi}$ is set, we are able to provide the value of the constant $G_0$ in Eq.(\ref{gform}) by means of Eq.(\ref{omegas}}.

In order to get a direct comparison with respect to a flat $\Lambda$CDM model, we now recast Eq.(\ref{Hform}) as
\textitb{\begin{align}\label{Hform-bis}
H_{\textrm{sr}}(z) = \hz(z)\; \sqrt{\omsn (1+z)^{3}+(1-\omsn)}\;,
\end{align}
where we have introduced the identification subscript (sr)} and we see how the slow-rolling paradigm induces an effective time dependent Hubble constant having the form
\begin{equation}\label{Heff}
\hz(z) = \hp \sqrt{\frac{\ompl (1+z)^{3}+(1-\ompl)+\Omega_{0\phi} (1+z)^{-3\beta}}{\omsn (1+z)^{3}+(1-\omsn)}}\; .
\end{equation}

This function may decays (depending on the model parameter $\beta$) as the redshift increases and, in what follows, we use this profile to account for the Hubble tension. The equation above is indeed the physical signature for the considered model, which has to be compared with other theoretical formulations and with data sets. We remark that, we have constructed a consistent picture in which the potential term of the scalar field is compatible with a slow-rolling phase (see below), providing the vacuum energy contribution to the present Universe. Furthermore, the deviation of the potential term from an exact constant energy density ensures that its kinetic contribution can rescale the value of the apparent Einstein constant, so that a varying Hubble constant with the redshift emerges.

\section{The Data sample and the fitting}\label{sec4}
To determine the model parameter $\beta$, let us now follow the data comparison described in \cite{apj-powerlaw,galaxies10010024}. For the sake of clarity \textitb{(here and in the following $H$ is in units: km s$^{-1}$ Mpc$^{-1}$)}, we specify that we set \cite{Planck:2018vyg,apj-powerlaw} 
\begin{align}
\hp=67.4\;,\qquad \ompl=0.315\;,\label{hplanck}\\
\hsn=73.5\;,\qquad \omsn=0.298\;\label{hsn},
\end{align}
for flat $\Lambda$CDM models. Specifically, we use the Pantheon sample data release \cite{Planck:2018vyg}, which is a compilation of 1048 SNe Ia taken from different surveys. In this analysis, we bin the data in redshifts so that the sample is equi-populated and we choose to divide the sample in 40 bins. 
We use the observed distance modulus, $\mu_{obs}=m_{B}-M$, taken as tabulated from Pantheon sample which contains the apparent magnitude in the B-band corrected for statistical and systematic effects ($m_{B}$) and the absolute in the B-band for a fiducial SN Ia with the stretch and color corrections ($M$).
Considering the color and stretch population models for SNe Ia, in this approach, similarly to the previous ones, we average the distance moduli given by the \cite{2010A&A...523A...7G} (G2010) and \cite{2011A&A...529L...4C} (C2011) models.
We then compare the $\mu_{obs}$ and the theoretical distance modulus that reads as follows:
\begin{equation}
\mu_{th}=5\hspace{0.5ex}log_{10}\ d_L(z,H_0,...) +25\;.
\label{eq1}
\end{equation}
The distance is usually expressed in Mpc \textitb{and it is constructed via a $\Lambda$CDM model}. Furthermore, we also consider the correction of the luminosity distance keeping into account the peculiar velocities of the host galaxies which contain the SNe Ia. To perform our analysis, we define the $\chi^2$ for SNe:
\begin{equation}
\label{eq9}
\chi^2_{SN}=\Delta\mu^{T}\cdot\mathcal{C}^{-1}\cdot\Delta\mu\;.
\end{equation}
Here $\Delta\mu=\mu_{obs}-\mu_{th}$, and $\mathcal{C}$ denotes the $1048 \times 1048$ \textit{covariance matrix}, given by \cite{2018ApJ...859..101S}. From the analysis of these bins, we obtain the values of $H_0$ in each bin by varying $H_0$ only and using $\omsn$. The choice of using 40 bins is dictated by two factors: the first is that we need enough number of data points to fit the model. The second is that although it would be beneficial to have a larger number of data points for fitting the theoretical model, however increasing the bins would result in a less precise fitting and thus a larger values of uncertainties on $H_0$. This would, hence, again result in less constraining and predictive power for the theoretical model. Indeed, this choice of binning allows us to have roughly 50 SNe Ia in each bin. We also note that the trend found is independent from the initial choice of $H_0$. This is due to the fact that we use the distance moduli in the Pantheon sample but subtracted of the value of the absolute magnitude, $M$, which corresponds to a given value of $H_0$. For example, we here consider the value of $M=-19.245$ as discussed in \citep{apj-powerlaw}. Thus, for different values of the starting $H_0$ we should change the values of $M$ correspondingly and this will allow a mere translation in the initial intercept and not a change of the slope itself. In fact the data of the Pantheon sample have a predefined $M$ fixed at $-19.35$, which we have changed to consider the starting point of $H_0=73.5$. 

\textitb{Now, we discuss the physical meaning of the function $\hz(z)$ in the proposed scenario. Clearly $H_0$ is a constant quantity by definition, so the running of the Hubble constant with the redshift must be interpreted as an effective behavior. In other words, we could set $\hz(z)\equiv H_0 h(z)$, where the function $h(z)$ admits a natural definition from Eqs.(\ref{Hform-bis}) and (\ref{Heff}). It is important to stress that, while the SHOES collaboration provides only the value $H_0$ from the SN Ia data, in \cite{apj-powerlaw,galaxies10010024} it was argued that the Pantheon sample, once decomposed in
equipopulated redshift intervals, allows to determine a
slow decaying of the value $H_0$ as the bin redshift increases. If the
Hubble tension admits a solution in terms of new physics, as we state here,
then the effect should continuously affect the observations and then the
existence of an effective $\hz$ has to be considered an expected features of data from different redshift domains. We conclude this digression by observing that, we use $\hsn$ as provided in \cite{apj-powerlaw} (see Eq.(\ref{hsn})) in order to fix the best value for the $\beta$ parameter, but the comparison between the Hubble parameter of our model and that one of the $\Lambda$CDM one is constructed according to the picture proposed in \cite{2021MNRAS.505.3866E}, i.e. we reconstruct for the bins of the Pantheon sample their profile as $H(z)$.}

Our model contains actually two free parameters, but the quantity $\Omega_{0\phi}$ is fixed in Eq.(\ref{omegaphi}) when requiring $H_0=\hsn$. Using the values introduced above, we obtain
\begin{align}
\Omega_{0\phi}=0.189\;.  
\end{align}
Concerning the parameter $\beta$, we perform a fitting procedure of the expression $\hz$ in Eq.(\ref{Heff}) by means of the 40 bin distribution of the $H_0$ values described above. The best fit is found for 
\begin{align}\label{parfit}
\beta=-0.258\pm 0.026\;.
\end{align}
\textitb{This value is obtained minimizing the residual $\chi^2$, routinely implementing errors on $H_0$ data.}
% We have also computed the probability that this value of the fit is drawn by chance and we have obtained a p-value = $\mathcal{O}(10^{-12})$. As usual, the threshold value is put as $0.05$, so any p-value< 5\% can guarantee us that the results are not drawn by chance. This clearly show that our model has the predictive power on the data in relation to the $\beta$ parameter.
\begin{figure}[ht!]
\centering
\includegraphics[width=8.3cm]{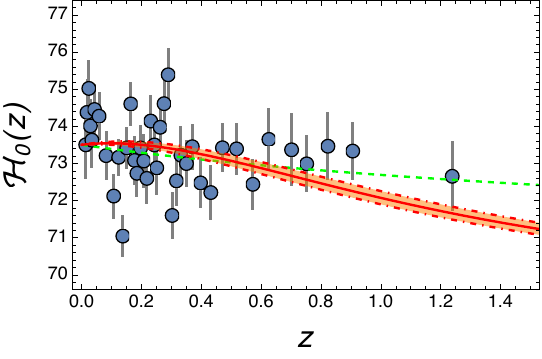}
\caption{\textitb{Plot of $\hz(z)$ from Eq.(\ref{Heff}) with $\beta=-0.258\pm0.026$ (red lines, with the errors)} and of the behavior $\hsn(1+z)^{-0.016}$ (green dashed line). Bullets represents $H_0$ data from \cite{apj-powerlaw} with the corresponding error bars. (For interpretation of the references to color in this figure legend, the reader is referred to the web version of this article.)
\label{figH0}}
\end{figure}
In \cite{apj-powerlaw}, it has been addressed a power-law profile as $\hz=\hsn (1+z)^{-\alpha}$ with the following best fit: $\alpha=0.016\pm0.009$, and in Fig.\ref{figH0} we show the data points and the fit of our slow-rolling model in red, together with the power-law behavior (green dashed line). %Although, the p-value of our fit is much smaller than the corresponding one for the power-law, i.e. our $\beta$ parameter appears better constrained by the data, however the global fit morphology does not allow to prefer one model to the other one in its capability to reproduce data.

\textitb{Let us now analyze the statistical relevance of our model in comparison with respect to the power-law behavior described above and also to the $\Lambda$CDM model with a constant $\hz=73.5$ (coherently with the binned data analysis). The values for $\chi^2$ test are the following (with obvious notation): $\chi^2_{\textrm{sr}}=0.452$ (dof=38), $\chi^2_{\textrm{pl}}=0.400$ (dof=38), $\chi^2_{\Lambda}=0.430$ (dof=39). Due to the high density of data at small $z$ and the significant number of dof, the obtained (inverse) p-values are almost $1$ for the three cases, denoting that the null hypothesis (the predictive power on data) is supported for all the models (the critical value is standard set as 0.05). Actually, the ordering $\chi^2_{\textrm{pl}}<\chi^2_{\Lambda}$ provides the evidence of the better significance of the power-law model in fitting the binned data, as outlined in \cite{apj-powerlaw}. At the same time, we stress how, nonetheless the slow-rolling model has the highest $\chi^2$ value when constraining $\beta$, the physical relevance of the obtained behavior relies on its capability to rapidly converge towards the proper Plank flat $\Lambda$CDM profile of $H(z)$, reducing the tension, as described in the following.}

\begin{figure}[ht!]
\centering
\includegraphics[width=8.3cm]{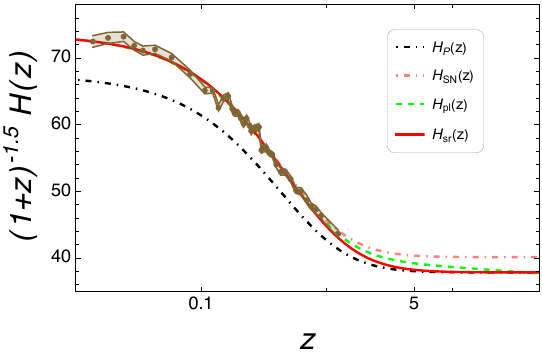}
\caption{Linear log-plot of normalized $H_{\textrm{sr}}(z)$ in Eqs.(\ref{Hform-bis}) with $\beta=-0.258$ (red line). Green-dashed line: $H_{\textrm{pl}}(z)$. Dotted-dashed lines represent: $H_{\textrm{SN}}(z)$ (pink) and $H_{\textrm{P}}(z)$ (black). Data points are properly scaled from Fig.\ref{figH0}.(For interpretation of the references to color in this figure legend, the reader is referred to the web version of this article.)
\label{figHz}}
\end{figure}

\textitb{In order to compare our results with respect the flat $\Lambda$CDM profiles, we introduce the following quantities
\begin{align}
&H_{\textrm{P}}(z)=\hp(\ompl (1+z)^{3}+(1-\ompl))^{0.5}\;,\\
&H_{\textrm{SN}}(z)=\hsn(\omsn (1+z)^{3}+(1-\omsn))^{0.5}\;,\\
&H_{\textrm{pl}}(z)=\hsn(1+z)^{-0.016}(\omsn (1+z)^{3}+(1-\omsn))^{0.5}\;,
\end{align}
with the definition in Eqs.(\ref{hplanck}) and (\ref{hsn}).} In Fig.\ref{figHz} we plot, also outlining the scaled data points from Fig.\ref{figH0}, $H_{\textrm{sr}}(z)$ form Eq.(\ref{Hform-bis}) of our slow-rolling model (red line) together with $H_{\textrm{pl}}(z)$ obtained by adopting the power law model in \cite{apj-powerlaw}, and also the two flat $\Lambda$CDM profiles: $H_{\textrm{P}}(z)$ and $H_{\textrm{SN}}(z)$ (as indicated in the figure).
It is evident how the contribution of the scalar field has a basic role at low $z$ values, allowing that $\hsn$ is recovered for $z\to0$. Furthermore, for $z\gtrsim5$, $H(z)$ clearly overlap the flat $\Lambda$CDM model associated to the Planck data (a similar behavior is obtained by the power-law model). \textitb{In particular, the obtained value of $\beta$ guarantees that the asymptotic ($z=1100$) behavior of $H(z)$ match $H_{\textrm{pl}}(z)$ with an error of $\mathcal{O}(10^{-7})$.}

The essential equivalence of the model here proposed and the power-law one
can be interpreted by observing that the modified gravity model adopted
to account for the second case  \cite{apj-powerlaw,galaxies10010024,schiavone_mnras} can also be
reduced to a scalar-tensor formulation. In fact, in the so-called
``Einstein framework'', the metric $f(R)$-gravity can be stated as standard
General Relativity minimally coupled to a scalar field, which
self-interaction potential is fixed by the specific modified
model of gravity. In \cite{schiavone_mnras}, the dynamics has been described in
the so-called ``Jordan frame'' in order to avoid the conformal rescaling of
the metric (necessary to pass to the Einstein framework) but, overall, to interpret the Hubble tension via an effective Einstein constant due to the non-minimally coupled scalar field. Anyway, both in the present dynamical scheme and in the idea proposed in \cite{schiavone_mnras}, the underlying mechanism is a slow-rolling
phase of a scalar component of the Universe. Clearly, modifying the
Einstein gravity has a number of additional predicted effects, with
respect to the solution of the Hubble tension, which the present model
certainly avoids: the energy density of the scalar matter, here adopted,
only affects the dynamics on the cosmological scales without other
significant implications on the early or shorter scale physics of the
Universe. \textitb{The analysis above calls attention to be extended to the Pantheon+ sample \cite{Brout:2022vxf}, especially because the enrichment of the SN Ia population could reliably improve the statistics of the binned analysis. Clearly, only after a new set up of $\mathcal{H}_0(z)$ from a suitable binned analysis, it would be possible to repeat the fitting procedure above and this requires a significant data analysis effort, which is of clear interest for future investigations, but exceeds the scope of the present modeling proposal for the Hubble tension.}
\begin{figure}[ht!]
\centering
\includegraphics[width=8.3cm]{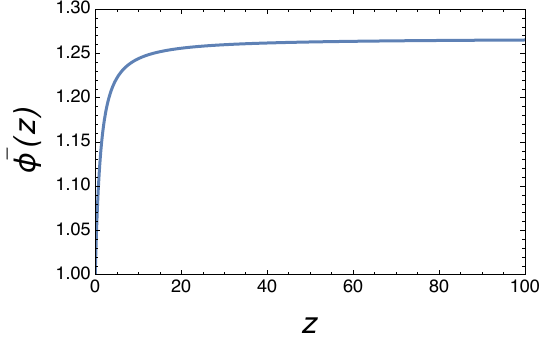}
\caption{Profile of $\bar{\phi}(z)$ evolved using Eq.(\ref{phievo}) by assuming $\bar{\phi}_0=1$. We have set $H_*=\hp$, $\Omega_{0\phi}=0.189$, $\beta=-0.258$ and the corresponding $H(z)$ expression in Eq.(\ref{Hform-bis}).
\label{figphiz}}
\end{figure}

\begin{figure}[ht!]
\centering
\includegraphics[width=8.3cm]{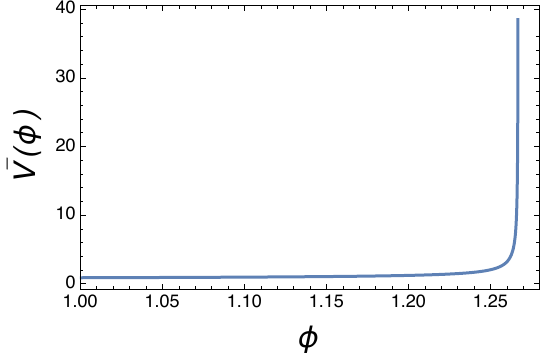}
\caption{Plot of $\bar{V}(\phi)$ from Eq.(\ref{ioma4}).
\label{figVphi}}
\end{figure}

We complete our analysis by showing, in Fig.\ref{figphiz}, the behavior of the normalized scalar field $\bar{\phi}$ evolved using Eq.(\ref{phievo}). The obtained profile starts at the arbitrarily set initial value $\bar{\phi}_0=1$ and frozen into an asymptote for $z\gtrsim60$. In Fig.\ref{figVphi}, we also plot the normalized potential $\bar{V}(\phi)=V\,\chi/3\hp^2$ from Eq.(\ref{ioma4}) and it is important to stress that the slow rolling conditions expressed again in Eq.(\ref{ioma4}) are properly satisfied for such a profile. In particular, we obtain Max$(|\dot{\phi}/2V(\phi)|)=0.125$ and Max$(|\ddot{\phi}/3H\dot{\phi}|)=0.13$. 

Finally, a self-interacting scalar field is always interpretable as a non-constant $w_{\phi}$-parameter of the equation of state, namely
\begin{equation}
w_{\phi} = \frac{\dot{\phi}^2/2 - V}{\dot{\phi}^2/2 + V}\, ,
\label{ioma18}
\end{equation}
which can be evaluated using Eq.(\ref{ioma8}) and the potential profile discussed above. It is evident from the the figure that this effective equation of state parameter is always associated to a dark energy contribution (it never becomes a phantom matter contribution) and, for $z\to0$, it correctly approaches a value very close to -1 ($w_\phi(0)=-0.95$).
\begin{figure}[ht!]
\centering
\includegraphics[width=8.3cm]{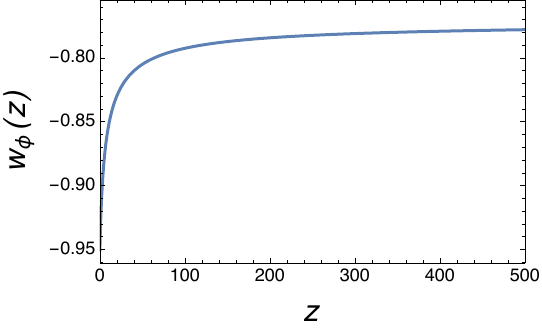}
\caption{Profile of $w_\phi(z)$  from Eq.(\ref{ioma18}).
\label{figwz}}
\end{figure}

\textitb{
Let us now phenomenologically discuss our model in the context of other well established paradigms. It is well-known \cite{DiValentino:2020naf,2022JHEAp..34...49A,2024PhRvD.109b3527A} that, if we consider an Hubble parameter in the form of a $w$CDM model specified for the SN Ia parameters ($\hsn$, $\omsn$), 
% \begin{equation}
% H_{\textrm{w}}(z) = \hsn\sqrt{\omsn(1+z)^3 + 
% (1 - \omsn)(1+z)^{3(1+w)}}\, , 
% \label{pm1}
% \end{equation}
% \begin{equation}
% H_{\textrm{w}}(z) = \hp\sqrt{\ompl(1+z)^3 + (c_1 - \ompl)(1+z)^{3(1+w)}}\, , 
% \label{pm1}
% \end{equation}
then, at small $z$, the profile can be reconciled with $H_{\textrm{P}}(z)$ if $w <-1 $, i.e. if a phantom matter contribution is introduced in the late Universe, thus alleviating the tension. On the contrary, taking $w >-1$, i.e. a physical quintessence contribution, the match mentioned above will be unflavoured, thus exacerbating the tension. This is true also for the $w_0 w_a$CDM model for what concerns variations of $w_0$.
% \begin{figure}[ht!]
% \centering
% \includegraphics[width=8.3cm]{fig_Hw}
% \caption{Linear log-plot of normalized $H_{\textrm{w}}(z)$ in Eqs.(\ref{pm1}) with $w-1.35$ (blue line) and $w-0.85$ (cyan line). Dotted-dashed lines represent: $H_{\textrm{SN}}(z)$ (pink) and $H_{\textrm{P}}(z)$ (black). (color online)
% \label{figHw}}
% \end{figure}

This consideration could appear in contradiction with the capability of our model to account for the Hubble tension, since its effective $w_\phi(z)$ remains always greater than $-1$, as show in Fig.\ref{figwz}. However, it is important to stress that our paradigm is not that expressed in a $w$CDM model or in a $w_0 w_a$CDM model, but a different one closely resembling the analysis presented in \cite{2021MNRAS.505.3866E}. In fact, in Eq.(\ref{Hform}), $H_*$ is not $H_0$ (which, to solve the Hubble tension, must be identified with the SN Ia value), but it stands equal to the Planck value, as specified in Eq.(\ref{H0expression}). In particular, the obtained form of $w_\phi(z)$ can be expressed (for $z\lesssim1$) as $w_\phi(z)=w_0+w_a z (1+z)^{-1}$ with $w_0=-0.95$ and $w_a=0.045$. Using this formalism, Eqs.(\ref{Hform}) (with $H_*=\hp$) rewrites
\begin{align}
H_{\textrm{sr}}(z) &= \hp\sqrt{\ompl(1+z)^3 + (c_1 - \ompl)f(z)}\, , \label{pm1}\\
f(z)&=(1+z)^{3(1+w_0+w_a)}e^{-3w_a z(1+z)^{-1}}\;,
\end{align}
where $c_1=\hsn^2/\hp^2$ to take account of Eq.(\ref{omegaphi}). As a result, the comparison with a standard $w_0 w_a$CDM model cannot be directly implemented and, hence, also a mapping of the $w_\phi$. In fact, we are able, by the present model, to address the Hubble tension problem because it contains an additional matter source (expressed by $c_1$ in Eq.(\ref{pm1})). Summarizing, a quintessence-like source can allow to account for the Hubble tension if it is regarded as an additional contribution not considered in the Planck measurement for the $\Lambda$CDM model, which has the important role to replace, in $z=0$, $\hp$ by $\hsn$, while at higher $z$-values it is progressively less relevant in fixing the profile of the Hubble parameter.

Finally, it is worth noting that, in order our model exactly fulfills the qualitative proposal in \cite{2021MNRAS.505.3866E}, it would be necessary to deal with a positive $\beta$ parameter, but this is ruled out by the fitting procedure on the binned SN Ia Pantheon analysis. This consideration would suggest that, to fit the low-redshift data, we have to deal with a Hubble parameter worse fitted by the Planck $\Lambda$CDM-analysis. We want to stress that, nonetheless this aspect, the integrated luminosity distance of our model differs (with the errors) of $\mathcal{O}(10^{-3})$ with respect to the Planck $\Lambda$CDM model. Deepening this question is out of the scope of the present study, but call attention for further investigations, based on different data samples.
}

\section{Concluding remarks}
We constructed a cosmological model with the aim to provide an explanation
for the Hubble tension problem, based on the introduction of scalar matter
in a slow-rolling configuration.
The merit of the scenario we proposed consists in the possibility to
account for the Universe vacuum energy via a constant term in the
self-interaction potential, while the scalar field kinetic energy is
responsible for a slow apparent variation of the Hubble constant, due to a
suitable rescaling of the Einstein constant, when a specific dynamical scheme is set up. In fact, we added to the Friedmann equation a specific
relation between the kinetic and potential term of the scalar field.
Together with the Klein-Gordon equation, we then
obtain three dynamical equations, in terms of the three unknowns
$H(z)$, $\phi(z)$ and $V(z)$
(from the latter two quantities, we calculated the functional form of $V(\phi
)$).

The Hubble function is then restated in terms of an effective varying
Hubble constant $\mathcal{H}_0(z)$, which contains two free parameters. One
of these has been fixed to reproduce the ratio between $\hsn$ and $\hp$, while the other free parameter was determined by a fitting analysis, via the 40 bin data of the SN Ia Pantheon sample \cite{apj-powerlaw}. We also compared the
quality of our fit with the one yielding the power-law behavior
$\hsn(1+z)^{-\alpha}$ and the profile of
the emerging best fit is essentially comparable in the two cases.

Then, we used the obtained values of the two free parameters to
plot the function $H(z)$, which, for $z\to0$ approaches the SN Ia data predictions for the Hubble constant while, for higher $z$-values,
it correctly matches the behavior of the corresponding curve specified by the
Planck detected parameters in the flat $\Lambda$CDM scenario. Actually, our $H(z)$ curve overlaps the proper profile already
for $z\gtrsim 5$, so offering an intriguing marker to discriminate between
low-redshift and high-redshift sources, when the Hubble constant is
determined according to a similar feature discussed in \citep{deangelis-fr-mnras}. In addition, variation of the Hubble constant from the central value obtained by the SNe Ia are visible with the use of Gamma Ray Bursts \citep{Dainotti2023ApJ...951...63D} and Quasars \cite{Dainotti2023ApJ...951...63D, Bargiacchi2023MNRAS.521.3909B,Lenart,Dainotti2023mnras} obtained with the use of new statistical assumptions discussed in \citep{Dainotti:2023ebr}.

The very significant novelty of the present study is, on the one hand, the
possibility to solve the Hubble tension problem via a simple scalar
matter model in a slow-rolling configuration and, on the other hand, that
our function $H(z)$ has been determined by using the whole information
contained in the SN Ia data (the 40 bins analysis of the Pantheon sample, as depicted in Fig.\ref{figHz})
and so avoiding the compression of the SN Ia set up to the determination
of the single value of $H(z=0)$, see the discussion in \cite{2021MNRAS.505.3866E}.

%\bibliographystyle{elsarticle-num.bst}
%\bibliography{biblio_astro}

\end{document}